\documentclass[aps,prl, twocolumn]{revtex4}%
\usepackage{amsfonts}
\usepackage[T2A]{fontenc}
\usepackage[cp1251]{inputenc}
\usepackage{amsmath}
\usepackage{amssymb}
\usepackage[russian,english]{babel}
\usepackage{graphicx}%
\begin{document}

\title{On the theory of magnetic field dependence of heat conductivity in dielectric
in isotropic model}
\author{L.A.Maksimov}
\author{T.V.Khabarova}
\email{frau_sych@mail.ru} \affiliation{Kurchatov Institute, Moscow
123182, Russia}

\begin{abstract}
Phonon polarization in a magnetic field is analyzed in isotropic model. It is
shown, that at presence of spin-phonon interaction phonon possess circular
polari-zation which causes the appearance of heat flux component perpendicular
both to temperature gradient and magnetic field.

\end{abstract}


\maketitle
\bigskip

Recently \cite{strohm}, \cite{inushtald} a novel phenomenon was found. It is
the detection in dielectric of a heat flow perpendicular both to magnetic
field$\ B$ and temperature gradient $\nabla T$ -- so called Phonon Hall Effect
(PHE). Such flow is caused by spin-phonon interaction (SPI) of phonons and
paramagnetic ions $Tb^{+3}$. The theory of PHE was considered in works
\cite{sheng} and \cite{KM} (the critic of work \cite{sheng} see in \cite{KM}).

However not only SPI leads to magnetic field dependence of dielectric heat
conductivity. As it is known, in molecular gases magnetic field affects the
transport phenomena because of the molecular rotary moments $\mathbf{M}$,
precessed in a magnetic field, dependence of collision cross-section
(Senftleben-Beenakker Effect \cite{beenakker}).

We expect, that the phenomenon similar to PHE in $Tb_{3}Ga_{5}O_{12}$ can be
found out in molecular crystals where is a component, consisting of the
molecules with rotary degrees of freedom (RDF). Typical representatives of
such substances are criocrystals \cite{crio}. To magnetic field dependence of
heat conductivity can lead a RDF-dependent phonon scattering on molecules. The
suitable theory is easy to construct via generalizing Senftleben-Beenakker
theory on a case of heat transfer by phonons. In approach where RDF are
considered quasi-classically such theory is considered in \cite{habar}. In
this paper we have another approach to this problem, and we show that in
crystals with RDF there is a mechanism close to SPI in ionic crystals which
leads to PHE through renormalization of acoustic waves because of lattice
vibrations interaction with molecular rotation. From symmetry considerations
describing such interaction Hamiltonian can be written down in similar to SPI
form:%
\begin{equation}
H_{1}=-g\sum_{n}\left(  \mathbf{M}_{n},[\mathbf{u}_{n}\times\mathbf{p}%
_{n}]\right)  .\label{1}%
\end{equation}
Here $\mathbf{M}_{n}$ is the rotary moment of the molecule in a cell $n$,
$[\mathbf{u}_{n}\times\mathbf{p}_{n}]$ -- total orbital moment of oscillating
atoms in a cell. Speaking generally, molecular crystals have a complex
structure containing of many particles in an elementary cell. But at low
temperatures heat is transferred by long-wave elastic waves, when all
particles of a cell oscillate with the same amplitude $\mathbf{u}_{n}$ and
velocity $\mathbf{v}_{n}$. Therefore instead of a multicomponent crystal it is
possible to consider a crystal with one atom in a cell with mass $m$, equal to
total mass of particles in a cell and one rotary moment. The value of $g$ in
case of SPI was estimated in numerous works on spin-lattice relaxation
\cite{KM} and \cite{works}. In case of acoustic vibrations interaction with
rotary degrees of freedom factor $g$ we shall consider as small
phenomenological parameter having the same order as for SPI when the molecule
has uncompensated electronic moment, as in $O_{2}$. We shall use system of
units where $k_{B}=1,\ \hbar=1.$ For short, interaction (\ref{1}) shall be
also called SPI in the case of crystals with rotary degrees of freedom. Almost
identical form of Hamiltonian describing phonon interaction with internal
degrees of freedom in ionic and molecular crystals (components of operator
$\mathbf{M}$ are replaced with Pauli matrices in ionic crystals with
quasi-doublet structure of the bottom levels) leads to qualitative similarity
of transverse thermal conductivity theories in both cases. But from the
physical point of view interaction mecha-nisms are far from each other.

Usual dielectric heat conductivity weakly depends on detailed properties of
acoustic vibrations. As it is shown in \cite{KM}, PHE is caused by occurrence
of phonon elliptic polarization at presence of a magnetic field and SPI. But
character of acoustic branches polarization essentially depends on symmetry
properties of a dynamic matrix. Unlike the general case considered in
\cite{KM}\ , where all three vibration modes are nondegenerate almost at all
directions of wave vector $\mathbf{k}$, in present paper the model of
vibrations in isotropic medium where transverse acoustic modes are degenerate
will be considered at all $\mathbf{k}$. Though this model is realized only
when atoms with a large interaction radius interact, in this model dynamic
matrix has rather simple form. In isotropic medium model we shall find a heat
flow in a $\left[  B\times\nabla T\right]  $ direction as for crystals with
rotary degrees of freedom, and for ionic crystals with quasi-doublet
structure. The first result is interesting as a prediction of PHE presence in
molecular crystals, and the second is correction of work \cite{sheng}.

In an external magnetic field the average value of the rotary moment
$\left\langle \mathbf{M}\right\rangle =\left\langle \mathbf{M}_{n}%
\right\rangle $ is nonzero, and crystal elastic vibrations are determined by
Hamiltonian renormalized taking into account (\ref{1}):%
\begin{equation}
H=\sum_{n}h_{n},\ \label{153}%
\end{equation}
where%
\begin{equation}
h_{n}=\frac{1}{2m}p_{n}^{2}-\frac{1}{2}\sum_{n^{\prime}}mD_{nn^{\prime}}%
^{ab}u_{n}^{a}u_{n^{\prime}}^{b}-g\left\langle \mathbf{M}\right\rangle \left[
\mathbf{u}_{n}\times\mathbf{p}_{n}\right]  .\label{154}%
\end{equation}
It is essential, that the effective coupling constant $g\left\langle
\mathbf{M}\right\rangle $ is proportional to magnetization of particles and it
disappears when someone switches off the magnetic field.

Using (\ref{154}) it is possible to write the quantum energy continuity
equation and to deduce the density of a heat flow formula (\cite{sheng},
\cite{KM}, \cite{hardy})%
\begin{equation}
j_{H}^{\gamma}=\frac{m}{2V}\sum_{nn^{\prime}}r^{\gamma}D_{nn^{\prime}}%
^{ab}u_{n}^{a}v_{n^{\prime}}^{b}.\label{2}%
\end{equation}
We emphasize, that in this expression term $v_{n}^{b}$\ is a velocity of this
ion, but not an impulse of an ion $p_{n}^{b}$ divided by mass
\begin{equation}
v_{n}^{a}=\partial_{t}u_{n}^{a}=\partial H/\partial p_{n}^{a}=p_{n}%
^{a}/m-e_{abc}g\left\langle M^{b}\right\rangle u_{n}^{c}.\label{77}%
\end{equation}
Let's turn in (\ref{2}) to impulse representation using%
\[
\sum_{r}r^{\gamma}D_{r}^{ab}\exp\left(  ikr\right)  =i\nabla_{k}^{\gamma}%
D_{k}^{ab}.
\]
We find%
\begin{equation}
j_{H}^{\gamma}=i\frac{m}{2V}\sum_{ksk^{\prime}s^{\prime}}\left(  \nabla
_{k}^{\gamma}D_{k}^{ab}\right)  u_{ks}^{a}v_{-ks^{\prime}}^{b}.\label{09}%
\end{equation}
($s$ is the mode number) Let's introduce decomposition of displacement vectors
and velocity vectors for normal vibration modes
\begin{equation}%
\begin{array}
[c]{c}%
u_{ks}^{a}=\sqrt{\frac{1}{2m\omega_{ks}}}(e_{ks}^{a}a_{ks}+e_{-ks}^{a\ast
}a_{-ks}^{+}),\\
v_{ks}^{a}=\left(  -i\omega_{ks}\right)  \sqrt{\frac{1}{2m\omega_{ks}}}%
(e_{ks}^{a}a_{ks}-e_{-ks}^{a\ast}a_{-ks}^{+}).
\end{array}
\label{22}%
\end{equation}
It is very important, that usual form has a velocity decomposition instead of
an impulse $\mathbf{p}_{i}=m\mathbf{v}_{i}-gm\left(  \mathbf{u}_{i}%
\times\mathbf{M}\right)  $ as authors of \cite{sheng} supposed. We substitute
$u$ and $v$ in (\ref{09}) for (\ref{22}) and average by state, which is
diagonal on phonon numbers. Then we discard anomalous averages $\left\langle
a_{ks}a_{-ks^{\prime}}\right\rangle $ and $\left\langle a_{-ks^{\prime}}%
^{+}a_{ks}^{+}\right\rangle $ and we come to heat flow in a $\left[
B\times\nabla T\right]  $ direction:%
\begin{multline}
\left\langle j_{H}^{y}\right\rangle =\frac{1}{4V}\sum_{kss^{\prime}}%
\left(\sqrt{\frac{\omega_{ks}}{\omega_{ks^{\prime}}}}+\sqrt{\frac{\omega
_{ks^{\prime}}}{\omega_{ks}}}\right)
\\
\times\left( \nabla_{k}^{y}D_{k}^{\alpha\beta }\right)
e_{ks}^{\alpha\ast}e_{ks^{\prime}}^{\beta}
\left\langle a_{ks}%
^{+}a_{ks^{\prime}}\right\rangle .\label{3}%
\end{multline}
In isotropic body PHE is characterized by one factor (as a Hall
constant). Without loss of generality we choose magnetization on
$Z$-axis, a temperature gradient -- on $X$, and a heat flow -- on
$Y$. Using this expression in \cite{KM} a PHE amount for acoustic
vibrations of a general view has been calculated. In present paper
we consider an elastic vibrations model in isotropic body where the
dynamic matrix in zeroth-order approximation $D_{k}^{ab}$ has a form
\begin{equation}
D_{k}^{ab}=c_{0}^{2}\delta^{ab}k^{2}+wk_{a}k_{b},\label{31}%
\end{equation}
with degenerate branches $\omega_{+,-}^{2}=\omega_{0}^{2}=c_{0}^{2}k^{2}$\ and
longitudinal branch $\omega_{\parallel}^{2}=\omega_{0}^{2}+\lambda
,\ \lambda=wk^{2}.$ Here the third independent term like $\delta^{ab}k_{a}%
^{2}$, which in a cubic crystal removes degeneration is discarded.

For isotropic body model considering orthogonality of polarization vectors
$e_{ks}^{a\ast}e_{ks^{\prime}}^{a}=\delta_{ss^{\prime}}$ we have
\begin{multline}
\left\langle j_{H}^{y}\right\rangle =\frac{1}{4V}\sum_{kss^{\prime}}%
(\sqrt{\frac{\omega_{ks}}{\omega_{ks^{\prime}}}}+\sqrt{\frac{\omega
_{ks^{\prime}}}{\omega_{ks}}})[\delta_{ss^{\prime}}2c_{0}^{2}k^{y}+
\\
+w(\left( k_{a}e_{ks}^{a\ast}\right)
e_{ks^{\prime}}^{y}+e_{ks}^{y\ast}\left(
k_{b}e_{ks^{\prime}}^{b}\right)  )]\left\langle a_{ks}^{+}a_{ks^{\prime}%
}\right\rangle .\label{4}%
\end{multline}
We see that (\ref{2}), (\ref{3}), (\ref{4}) do not contain factor
$g$ evidently. The heat flow magnitude changes only because of
renormalization of phonon spectrum and polarization. Everyone can
verify that diagonal terms in (\ref{4}) do not lead to heat flow in
$\left[  B\times\nabla T\right]  $ direction. PHE is described by
off-diagonal terms in (\ref{4}):
\begin{equation}
\left\langle j_{H}^{y}\right\rangle =\frac{w}{2V}\sum_{kss^{\prime}}%
(\sqrt{\frac{\omega_{ks}}{\omega_{ks^{\prime}}}}+\sqrt{\frac{\omega
_{ks^{\prime}}}{\omega_{ks}}})\operatorname{Re}[e_{ks}^{y\ast}\left(
\mathbf{k}\mathbf{e}_{ks^{\prime}}\right)  \left\langle a_{ks}^{+}%
a_{ks^{\prime}}\right\rangle ]\label{40}%
\end{equation}
Particularly, transverse modes correlated motion gives%
\begin{multline}
\left\langle j_{\perp}^{y}\right\rangle =\frac{w}{2V}\sum_{k}\operatorname{Re}%
\{e_{k-}^{y\ast}\left(  \mathbf{k}\mathbf{e}_{k+}\right)  )\left\langle
a_{k-}^{+}a_{k+}\right\rangle +
\\
+e_{k+}^{y\ast}\left(  \mathbf{k}\mathbf{e}%
_{k-}\right)  \left\langle a_{k+}^{+}a_{k-}\right\rangle \}.\label{5}%
\end{multline}

Expression (\ref{5}) essentially depends on phonon polarization vectors. First
of all it is necessary to establish their form in consideration of SPI.
Hamiltonian (\ref{154}) leads to the dispersion equation%
\begin{equation}
\omega_{ks}^{2}e_{ks}^{a}=\tilde{D}_{k}^{ab}e_{ks}^{b},\label{0157}%
\end{equation}
where%
\begin{align}
\tilde{D}_{k}^{ab}  &  =D_{k}^{ab}+iD_{1k}^{ab},\ \ D_{1k}^{ab}=e_{abc}%
G^{c},\ G^{c}=2\omega g\left\langle M^{c}\right\rangle \label{158}
\end{align}
\begin{align}
\tilde{D}_{k}^{ab}\left(  \mathbf{G}\right)   &  =\left(  \tilde{D}_{k}%
^{ba}\left(  -\mathbf{G}\right)  \right)  ^{\ast},\ \ \left(  \mathbf{e}%
_{ks}^{\ast}\mathbf{e}_{ks^{\prime}}\right)  =\delta_{ss^{\prime}}.\label{159}%
\end{align}

SPI contribution to dynamic matrix is an imaginary antisymmetric tensor. In
model (\ref{31}) dispersion equation takes a form
\begin{equation}
We^{a}=\lambda\hat{k}^{a}(\hat{k}e)+i\left[  e\times\mathbf{G}\right]
^{a}\ \label{03}%
\end{equation}
Here $W=\omega^{2}-\omega_{0}^{2}$, $\lambda=wk^{2}$, $\hat{k}=\mathbf{k}%
/k=(\sin\theta\cos\varphi,\sin\theta\sin\varphi,\cos\theta)$ and
$\mathbf{G}=(0,0,G)$, $\left(  \mathbf{G}\hat{k}\right)  =G\cos\theta$,
$Q=G\sin\theta$. Let's use orthonormal basis vectors%
\begin{align*}
&1)\hat{k},\ 2)\hat{m}=(\mathbf{G}-\left( \mathbf{G}\hat{k}\right)
\hat{k})/Q=\\
&=(-\cos\theta\cos\varphi,-\cos\theta\sin\varphi,\sin\theta),\\
&3)\hat{n}  =\left[  \hat{k}\times\hat{m}\right]  =Q^{-1}\left[ \hat
{k}\times G\right]  =(\sin\varphi,-\cos\varphi,0),
\end{align*}
From (\ref{03}) we find vector $\mathbf{e}$ components%
\begin{equation}
\mathbf{e}=\xi\lbrack\frac{Q}{\lambda-W}\hat{k}+\frac{\left(  \mathbf{G}%
\hat{k}\right)  }{W}\hat{m}-i\hat{n}]\label{07}%
\end{equation}
and eigenvalue equation%
\begin{equation}
\frac{Q^{2}W}{\lambda-W}+W^{2}-\left(  \mathbf{G}\hat{k}\right)
^{2}=0.\label{228}%
\end{equation}

Dispersion equation (\ref{0157}) determines polarization vectors accurate
within phase. We set phase of parameter $\xi$ equal to zero. Parameter
absolute value is set by normalization:%
\begin{equation}
\xi^{-2}=\left(  \frac{Q}{\lambda-W}\right)  ^{2}+\left(  \frac{\left(
\mathbf{G}\hat{k}\right)  }{W}\right)  ^{2}+1.\label{20}%
\end{equation}
At inversion ($\hat{k},\hat{m},\hat{n}\rightarrow-\hat{k},\hat{m},-\hat{n}$)
polarization vector (\ref{07}) changes its sign as a polar vector. Complexity
of vector (\ref{07}) means that it is elliptically polarized.

Expressions (\ref{07}), (\ref{228}), (\ref{20}) are correct with any $G$. Then
we assume phonon and internal degrees of freedom of ions (molecules)
interaction to be weak $G\ll\lambda$. In this case we get one longitudinal
mode $W_{\parallel}=\lambda,$ $\mathbf{e}=\hat{k}$ from (\ref{228}).

For transverse modes in zeroth-order approximation $W$ is equal to zero, and
in consideration of $G$ is equal to%
\begin{equation}
W_{\eta}=\frac{1}{2}(-\left(  Q^{2}/\lambda\right)  +\eta\sqrt{\left(
Q^{2}/\lambda\right)  ^{2}+4\left(  \mathbf{G}\hat{k}\right)  ^{2}},\ \eta
=\pm1
\end{equation}
and describes their splitting
\begin{equation}
\Delta=\omega_{+}-\omega_{-}=\frac{1}{\omega_{0}}\sqrt{\left(  Q^{2}%
/2\lambda\right)  ^{2}+\left(  \mathbf{G}\hat{k}\right)  ^{2}}.\label{18}%
\end{equation}
Splitting is minimal on equator ($\left\vert \cos\theta\right\vert
<G/\lambda,\ \Delta=Q^{2}/\left(  2\lambda\omega_{0}\right)  $). For other
$\hat{k}$ directions we have
\begin{equation}
W_{\eta}=\eta\left\vert \mathbf{G}\hat{k}\right\vert ,\ \Delta=G\left\vert
\cos\theta\right\vert /\omega_{0}.\label{200}%
\end{equation}

Real and imaginary parts of each $\mathbf{e}_{\eta}$ in zeroth-order
approximation mutually perpendicular end equal to each other.
\begin{equation}
\mathbf{e}_{\eta}\simeq\frac{1}{\sqrt{2}}[\eta\left(  sign\cos\theta\right)
\hat{m}-i\hat{n}].\label{19}%
\end{equation}
This means transverse phonons have circular polarization \cite{sheng}.

Polarization vector (\ref{19}) direction changes stepwise crossing equator.
However, if we change transverse mode numeration and write $W_{\eta}%
=\eta\mathbf{G}\hat{k}$ instead of (\ref{200}) then projection of (\ref{07})
on the $\hat{m}$-axis will be constant. It always happens when crossing
levels. The magnitude of (\ref{5}) does not depend on numeration.

In linear approximation we find
\begin{equation}
\left(  \mathbf{e}_{\eta}\hat{k}\right)  =\frac{\xi_{\eta}Q}{\lambda}\left(
1+\frac{W_{\eta}}{\lambda}\right)  .\label{13}%
\end{equation}
We also need following expressions:
\begin{equation}
\xi_{\eta}\simeq\frac{1}{\sqrt{2}}(1-\frac{Q^{2}}{4\lambda W_{\eta}%
}),\label{28}%
\end{equation}%
\begin{multline}
e_{\eta}^{y}=\xi_{\eta}[\frac{Q}{\lambda-W_{\eta}}\left(  \sin\theta
\sin\varphi\right)  +\frac{\left(  \mathbf{G}\hat{k}\right)  }{W_{\eta}%
}\left(  -\cos\theta\sin\varphi\right)  +
\\
+i\cos\varphi].\label{29}%
\end{multline}

Expression (\ref{5}) contains projection of polarization vector onto wave
vector direction $\left(  \mathbf{e}_{\eta}\hat{k}\right)  $ which is linear
on SPI. So, off-diagonal density matrix $\left\langle a_{ks}^{+}a_{ks^{\prime
}}\right\rangle $ is quite enough to calculate in zeroth-order approximation
on SPI.

As is known, the nonequilibrium part of density matrix (diagonal on modes) in
tau-approximation is equal%
\begin{multline}
f_{p}=\left\langle a_{p}^{+}a_{p}\right\rangle -N_{p}=-\frac{1}{\Omega_{pp}%
}\left(  \mathbf{c}_{p}\nabla\right)  N_{p}=
\\
=-\frac{1}{\Omega_{pp}T^{2}}%
N_{p}\left(  1+N_{p}\right)  \left(  \omega_{p}\mathbf{c}_{p}\right)  \nabla
T.\label{2008}%
\end{multline}
Here and afterwards, we use $p=ks$ for short, $1/\Omega_{pp}$ is a
$p$-phonon relaxation time. Furthermore, in obvious cases we shall
leave out an index $\mathbf{k}.$

The corresponding energy flow is parallel to $\nabla T$\ and leads to heat
conductivity coefficient%
\begin{equation}
\varkappa^{xx}\simeq T^{3}\left(  c\Omega\right)  ^{-1},\label{2003}%
\end{equation}
where $c$, $\Omega$\ -- average values of $c_{p},\Omega_{pp}.$

Off-diagonal part of density matrix can be expressed by $f_{p}$ constructing
generalized Boltzmann equation. Phonon scattering has many channels:
anharmonicity, resonant scattering with molecular multiplet structure
excitation, scattering with rotary degrees of freedom contribution, impurity
scattering. In all cases the answer has similar structure. In \cite{KM} it is
shown, that%
\begin{equation}
\left\langle a_{p}^{+}a_{q}\right\rangle =\frac{iJ_{pq}}{\left(  \omega
_{p}-\omega_{q}\right)  },
\end{equation}
where $J_{pq}$ is a Hermitian matrix which differs mainly from usual collision
integral in that there is product of the phonon scattering amplitudes instead
of a square of the scattering amplitude module (Fermi's golden rule). In
tau-approximation this expression has a form of%
\begin{equation}
\left\langle a_{p}^{+}a_{q}\right\rangle =i\frac{\Omega_{qp}\left(  \omega
_{p}\right)  f_{p}+\Omega_{qp}\left(  \omega_{q}\right)  f_{q}}{2\left(
\omega_{p}-\omega_{q}\right)  },\label{11}%
\end{equation}
where $\Omega_{qp}^{\ast}\left(  \omega_{p}\right)  =\Omega_{pq}\left(
\omega_{p}\right)  $ and $\left\langle a_{p}^{+}a_{q}\right\rangle ^{\ast
}=\left\langle a_{q}^{+}a_{p}\right\rangle $ are Hermitian.

The effective relaxation frequencies form depends on scattering mecha-nism. In
\cite{KM} corresponding expressions for resonant scattering and anharmonicity
are shown. In case of potential impurity scattering
\begin{equation}
\Omega_{qp}\left(  \omega_{p}\right)  =2\pi\frac{N_{im}}{N^{2}}\sum_{g}%
A_{qg}A_{gp}\delta\left(  \omega_{p}-\omega_{g}\right)  .\label{12}%
\end{equation}
It is essential, that the transition amplitude $A_{pq}=e_{p}^{a\ast}%
A_{pq}^{ab}e_{q}^{b}$ is proportional to polarization vectors. It guarantees
the absence of a stream (\ref{5}) dependence on general phases of phonon modes
(phases of parameter $\xi$ in (\ref{07})). The $A_{qg}$
polarization-dependence is automatically transferred on relaxation
frequencies$\ \Omega_{qp}=e_{q}^{a\ast}\Omega_{qp}^{ab}e_{p}^{b}$, where
tensor $\Omega_{qp}^{ab}=\left(  \Omega_{pq}^{ba}\right)  ^{\ast}$\ does not
depend on external polarization vectors. In isotropic body model we can
neglect mode splitting for transverse modes in numerator of (\ref{11})
\begin{equation}
\left\langle a_{k+}^{+}a_{k-}\right\rangle =i\Omega_{-+}\left(  \omega
_{0}\right)  \frac{f_{\perp}}{\Delta}\label{999}%
\end{equation}
When one of modes is longitudinal from (\ref{11}) we have%
\begin{equation}
\left\langle a_{3}^{+}a_{\eta}\right\rangle =\frac{i}{2\varepsilon}%
[\Omega_{3\eta}\left(  \omega_{3}\right)  f_{3}+\Omega_{\eta3}\left(
\omega_{0}\right)  f_{\perp}].\label{88}%
\end{equation}
where $\varepsilon=\omega_{3}-\omega_{0}=\sqrt{\omega_{0}^{2}+\lambda^{2}%
}-\omega_{0}$.

Let's use the formulas above for calculation of transverse heat conductivity
coefficient $\varkappa^{yx}.$ Using (\ref{999}) in (\ref{5}) we find
\begin{multline}
\left\langle j_{\perp}^{y}\right\rangle =\frac{w}{2V}\sum_{k}\operatorname{Re}%
\{-e_{k-}^{y\ast}\left(  \mathbf{k}\mathbf{e}_{k+}\right)  i\Omega_{-+}^{\ast
}\left(  \omega_{0}\right)  \frac{f_{\perp}}{\Delta}+
\\
+e_{k+}^{y\ast}\left( \mathbf{k}\mathbf{e}_{k-}\right)
i\Omega_{-+}\left(  \omega_{0}\right)
\frac{f_{\perp}}{\Delta}\}.\label{89}%
\end{multline}
In (\ref{2008}) function $f_{\perp}\ $is proportional to
$c^{x}\sim\cos \varphi,$ and nonzero result is given only by
imaginary part of (\ref{29}): $e_{k\eta}^{y\ast}=\left(  ...\right)
-i\xi_{\eta}\cos\varphi.$ From
(\ref{29}), (\ref{13}) and (\ref{89}) we have%
\begin{multline}
\left\langle j_{\perp}^{y}\right\rangle =\frac{w}{2V}\sum_{k}\frac{f_{\perp}%
}{\Delta}\frac{kQ}{\lambda}\xi_{-}\xi_{+}\cos\varphi
\\
\times\operatorname{Re}%
\{-\left(  1+\frac{W_{+}}{\lambda}\right)  \Omega_{-+}^{\ast}+
\left( 1+\frac{W_{-}}{\lambda}\right)  \Omega_{-+}\}.
\end{multline}
We see, that only second-order terms in (\ref{13}) give the
contribution to PHE. Using concrete form of parameters $\lambda,\
\Delta,\ Q,\ \xi_{\eta
},\ W_{\eta},\ G$ we find%
\begin{equation}
\left\langle j_{\perp}^{y}\right\rangle =-\frac{g\left\langle M\right\rangle
w}{V}\sum_{k}f_{\perp}\frac{\omega_{0}^{2}}{w^{2}k^{3}}\sin\theta\cos
\varphi\operatorname{Re}\Omega_{-+}.
\end{equation}
Then we integrate using (\ref{2008})%
\begin{multline}
\varkappa^{yx}=-\frac{g\left\langle M\right\rangle
w}{V}\sum_{k}\frac {\omega_{0}^{3}c}{w^{2}k^{3}}\left(  \sin\theta\cos\varphi\right)  ^{2}\frac{\operatorname{Re}%
\Omega_{-+}}{\Omega_{pp}T^{2}}
\\
\times N_{p}\left(  1+N_{p}\right)  .
\end{multline}
Consequently, accepting $w\simeq c^{2}\ $and $\left\langle \frac
{\operatorname{Re}\Omega_{-+}}{\Omega_{pp}}\right\rangle \simeq1,$
we find accurate within numerical coefficient
\begin{equation}
\varkappa^{yx}=-\frac{g\left\langle M\right\rangle
}{V}\sum_{k}\frac{c^{2}}{T^{2}}N_{p}\left(  1+N_{p}\right)
\simeq-\frac{g\left\langle
M\right\rangle T}{c}\label{144}%
\end{equation}
Dividing it on longitudinal heat conductivity coefficient $\varkappa
^{xx}\simeq T^{3}\left(  c\Omega\right)  ^{-1}$ we find a Hall angle
\[
\frac{\varkappa^{yx}}{\varkappa^{xx}}\simeq\left\langle \mathbf{M}%
\right\rangle \frac{g\Omega}{T^{2}}.
\]

The second channel of Hall heat conductivity is caused by occurrence in
temperature gradient presence a correlated transverse and longitudinal phonon
motion which is characterized by off-diagonal density matrix component
(\ref{88}):%
\begin{equation}
\left\langle a_{1}^{+}a_{3}\right\rangle =-\frac{i}{2\varepsilon}%
[\Omega_{3\eta}\left(  \omega_{0}\right)  f_{1}+\Omega_{\eta3}\left(
\omega_{3}\right)  f_{3}].\label{61}%
\end{equation}
Here $1=\left(  \mathbf{k},\perp\right)  ,\ 3=\left(  \mathbf{k}%
,\parallel\right)  $. This density matrix (and its Hermitian conjugation)
forms Hall heat flow (see (\ref{40}))%
\begin{multline}
\left\langle j_{H}^{y}\right\rangle
=\frac{w}{2V}\sum_{k}k(\sqrt{\frac
{\omega_{1}}{\omega_{3}}}+\sqrt{\frac{\omega_{3}}{\omega_{1}}}%
)\operatorname{Re}[e_{1}^{y\ast}\left(  \hat{k}\mathbf{e}_{3}\right)
\left\langle a_{1}^{+}a_{3}\right\rangle +
\\
+e_{3}^{y\ast}\left(  \hat{k}\vec
{e}_{1}\right)  \left\langle a_{3}^{+}a_{1}\right\rangle ]\label{62}%
\end{multline}
The matrix (\ref{88}) is proportional to (\ref{2008}))\ and
$\cos\varphi.$ Besides, $e_{k\eta}^{y\ast}=-i\xi_{\eta}\cos\varphi$
and $e_{3}^{y\ast}=\sin\theta\sin\varphi.$ So, only the first term
in square brackets gives a contribution to integral. In this case
$\left(  \hat {k}\mathbf{e}_{3}\right) \simeq1$ and integration
element in (\ref{62}) does not contain series expansion parameter of
SPI.
\begin{multline}
\left\langle j_{H}^{y}\right\rangle
=-\frac{w}{2V}\frac{1}{\sqrt{2}}\sum
_{k}(\sqrt{\frac{\omega_{1}}{\omega_{3}}}+\sqrt{\frac{\omega_{3}}{\omega_{1}}%
})\cos\varphi\frac{k}{2\varepsilon}(f_{1}\operatorname{Re}\Omega_{31}%
+
\\
+f_{3}\operatorname{Re}\Omega_{13}\left(  \omega_{3}\right)  )
\end{multline}
Using (\ref{2008}) and integrating we estimate longitudinal mode
contribution to Hall heat conductivity
\begin{equation}
\varkappa_{\parallel}^{yx}\sim\frac{T^{2}\operatorname{Re}\Omega_{31}}{\left(
c_{\parallel}-c_{0}\right)  \Omega_{pp}}\sim\frac{T^{2}}{c_{0}}\label{41}%
\end{equation}
In this case the Hall angle has an order of magnitude of
\begin{equation}
\frac{\varkappa_{\parallel}^{yx}}{\varkappa^{xx}}\sim\frac{\Omega}%
{T}\label{42}%
\end{equation}

In conclusion we compare the estimations of heat conductivity coefficients
caused by transverse and longitudinal mode correlation $\varkappa_{\parallel
}^{yx}$(\ref{41}), transverse modes $\varkappa_{\perp}^{yx}$(\ref{144}), and
estimation $\varkappa_{KM}^{yx}$ for ion crystal with nondegenerate modes
\cite{KM}
\begin{equation}
\varkappa_{\parallel}^{yx}\sim\frac{T^{2}}{c_{0}},\ \ \varkappa_{\perp}%
^{yx}\sim\left\langle M\right\rangle \frac{gT}{c_{0}},\ \ \varkappa_{KM}%
^{yx}\simeq\left\langle \sigma\right\rangle \frac{gT}{\bar{c}}.
\end{equation}
We see, that estimations for transverse modes in isotropic model and cubic
crystal coincide accurate within average polarization caused by rotary moment
substitu-tion for pseudospin polarization. It is essential, that in both cases
frequency de-pendence is absent as it should be for heat flow component in
$\left[  B\times\nabla T\right]  $. direction. The longitudinal mode
(\ref{41}) contribution has qualitative other form. It does not depend not
only on frequency but on coupling constant too, as it is in common Hall
Effect. At high temperatures ($T\gg g$) channel $\varkappa_{\parallel}^{yx} $
plays a main role. If there were crystals where even approximately there were
acoustic vibrations with ellipticity close to circular, then PHE experimental
observation would become rather simple problem.

The first theoretical work devoted to PHE \cite{sheng} contained a mistake at
using relation (\ref{77}) between particle velocity and their impulse at
presence of SPI. Nevertheless it is interesting to compare a $\varkappa^{yx}
$\ estimation in \cite{sheng} with our formulas (\ref{41}), (\ref{144}), as in
\cite{sheng} calculations were performed in isotropic model like in present
paper. If to not pay attention to numerical coefficients which are obvious
excess of accuracy, they re-ceived a result
\[
\varkappa_{sheng}^{yx}=\frac{gT}{c_{0}},
\]
which coincide with (\ref{144}) and main result of \cite{KM}. This concurrence
is caused not by of our theory equivalence, and only trivial consequence of
similar dimension of results and that all of them are received in linear
approach on $g$. So, it is necessary to emphasize, that the basic result of
the this work (and \cite{KM}) is not the estimation of $\varkappa^{yx}$, and
the establishment of that fact, that PHE is caused by joint action of two
equally important factors - 1) phonon elliptic polarization caused by SPI, and
2) induced by a temperature gradient the correlated movement of two phonon
modes with off-diagonal density matrix formation.

\end{document}